\documentclass[aps,prl,twocolumn,showpacs,amsmath,amsmath,amssymb]{revtex4-1}

\usepackage{amsmath,amssymb,graphicx,bm,psfrag,bbold,float,color}

\usepackage{physics,hyperref}

\usepackage[mathscr]{eucal}

\DeclareMathAlphabet{\mathpzc}{OT1}{pzc}{m}{it}

\begin{document}

\title{Feedback control of persistent-current oscillation based on the atomic-clock technique}

\author{Deshui Yu$^{1}$ and Rainer Dumke$^{1,2}$}

\address{\mbox{$^{1}$Centre for Quantum Technologies, National University of Singapore, 3 Science Drive 2, Singapore 117543, Singapore}}

\address{$^{2}$Division of Physics and Applied Physics, Nanyang Technological University, 21 Nanyang Link, Singapore 637371, Singapore}

\begin{abstract}
We propose a scheme of stabilizing the persistent-current Rabi oscillation based on the flux qubit-resonator-atom hybrid structure. The LC resonator weakly interacts with the flux qubit and maps the persistent-current Rabi oscillation onto the intraresonator electric field. This field is further coupled to a Rydberg-Rydberg transition of the $^{87}$Rb atom. The Rabi-frequency fluctuation of the flux qubit is deduced from measuring the atomic population and stabilized via feedback controlling the external flux bias. Our numerical simulation indicates that the feedback-control method can efficiently suppress the background fluctuations in the flux qubit, especially in the low-frequency limit. This technique may be extensively applicable to different types of superconducting circuits, paving a new way to long-term-coherence superconducting quantum information processing.
\end{abstract}


\maketitle

\textit{Introduction.} Hybridizing superconducting (SC) circuits and atoms is a promising idea for realizing quantum information processing, transfer and storage~\cite{RMP:Xiang2013,SciRep:Yu2016,QST:Yu2017,ProcSPIE:Hufnagel2017,PRL:Petrosyan2008,PRA:Patton2013,PRA:Yu2016-1,PRA:Yu2016-2}. Such a quantum scheme also provides a platform for investigating fundamental principles of ultrastrong interaction between electromagnetic fields and atoms~\cite{NatPhys:Niemczyk2010,NatPhys:Yoshihara2017,PRA:Yu2017}. Yet, despite all this, the rapid coherence decay of solid-state devices significantly restricts the practical implementation of these hybrid systems~\cite{RevMexFisS2011,NatCommun:Bernon2013,PRL:Weiss2015}.

\begin{figure}[b]
\includegraphics[width=8.0cm]{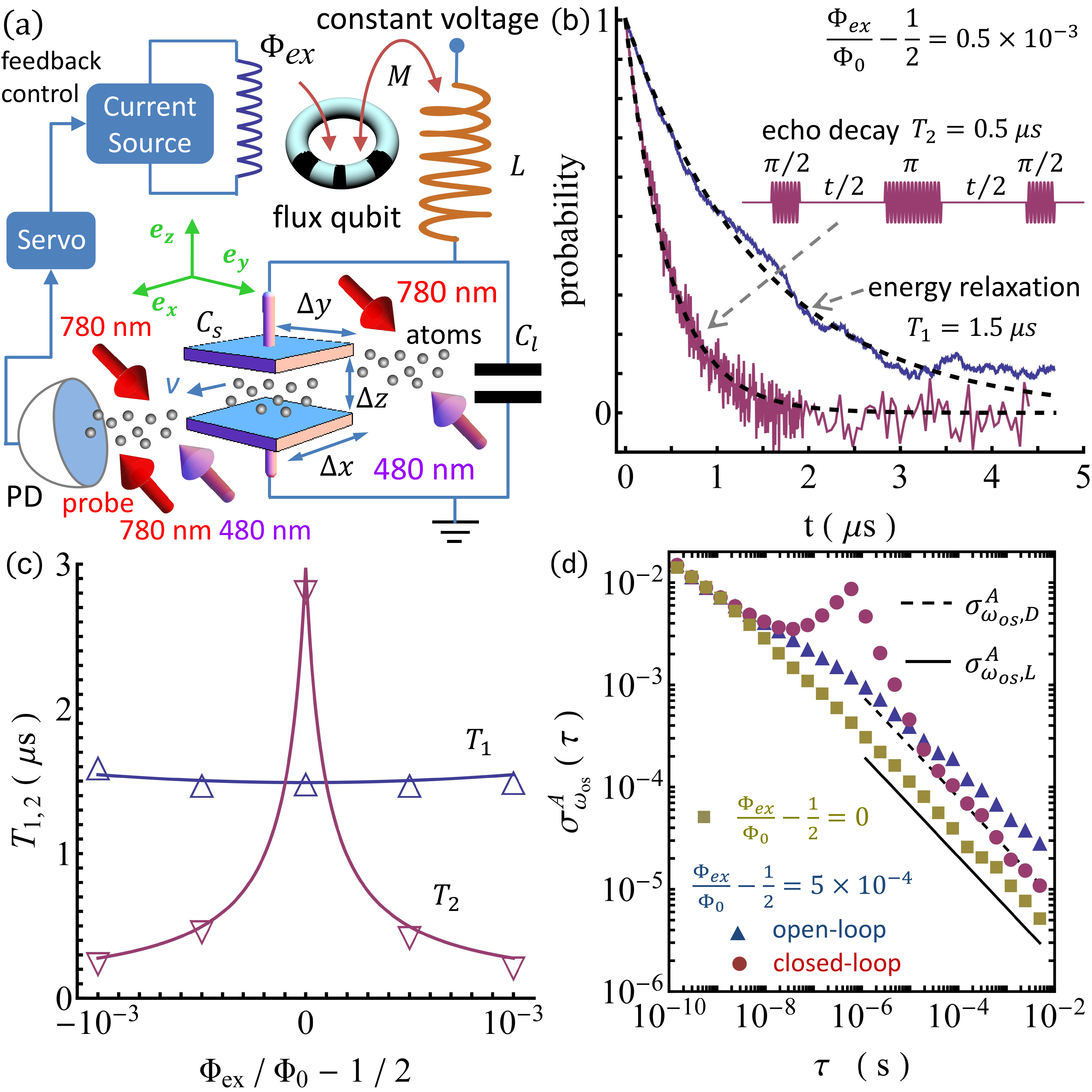}\\
\caption{(Color online) (a) Schematic diagram of flux-qubit-resonator-atom hybrid. The flux qubit is biased by $\Phi_{ex}$, generated by a constant-current loop, and inductively coupled to the series LC resonator. The atoms interact with the intraresonator field of $C_{s}$. The atomic population is measured via a photodetector by the resonance fluorescence. The measurement result is fed back into the constant-current loop via the servo. (b) Numerical simulation of energy relaxation and echo decay of flux qubit with $\frac{\Phi_{ex}}{\Phi_{0}}-\frac{1}{2}=5\times10^{-4}$ and the initial state $|1\rangle$. (c) Analytical and numerical results of energy-relaxation $T_{1}$ and dephasing $T_{2}$ times as a function of $\Phi_{ex}$. (d) Allan deviation $\sigma_{\omega_{os}}(\tau)$ of Rabi oscillator with different $\Phi_{ex}$ and feedback-control operation.}\label{Fig1}
\end{figure}

The atomic-clock technology has been proven as the most efficient tool to preserve the coherence of a local harmonic oscillator~\cite{Nature:Takamoto2005,NatPhys:Akatsuka2008,NatPhoton:Ushijima2015}. Employing the similar measurement to SC circuits potentially enhances their energy-relaxation and dephasing times. The decoherence mechanisms of various SC qubits have been systematically studied~\cite{PRL:Astafiev2004,PRB:Ithier2005,PRB:Lupascu2005,PRL:Anton2013}. On this basis, in~\cite{NJP:Yu2017} it has been theoretically demonstrated that the frequency fluctuations of a charge-qubit Rabi oscillation can be suppressed via the feedback-control method combined with probing the gate-voltage-bias noise. However, since it is not a direct measurement of the qubit-Rabi-oscillation-frequency fluctuations, the feedback-control efficiency could degrade. Additionally, in~\cite{NJP:Yu2017} the necessary condition, i.e., the efficient SC-qubit-atomic-reference coupling, is fulfilled by the direct electric-dipole interaction between the Rydberg atoms and the local electric field from the charge-qubit circuit. However, such a direct interface is extremely challenging in the flux-qubit-atom hybrid because of the weak magnetic-dipole interaction. To prevent these two limitations, it is necessary to establish a novel approach to directly measure the qubit-Rabi-oscillation-frequency fluctuations and generalize it to other types of SC circuits.

In this Letter, we explore the application of the feedback-control approach in maintaining the Rabi oscillation of the flux-qubit component in a hybrid structure. In this scheme, an LC resonator is inductively linked to the flux qubit and maps the persistent-current Rabi oscillation onto the intraresonator electric field which is further electrically coupled to a Rydberg-Rydberg transition of the $^{87}$Rb atom, resulting in a strong indirect flux-qubit-atomic-reference interface. The projection measurement of the atomic reference enables a direct probing of the Rabi-frequency fluctuations of the persistent current flowing in the flux-qubit loop. Feedback controlling the external flux bias enhances the long-term stability of the flux-qubit Rabi oscillator. Such a SC-qubit-resonator-atom stabilization scheme can be generalized to other types of SC qubits.

\textit{Flux qubit.} We consider a flux qubit~\cite{Science:Mooij1999} biased by an external magnetic flux $\Phi_{ex}$ which is produced by a constant-current source [Fig.~\ref{Fig1}(a)]. In the basis of clockwise $|L\rangle$ and counter-clockwise $|R\rangle$ persistent-current states, the flux-qubit Hamiltonian (without a flux drive) in the presence of energy fluctuations is written as $H/\hbar=H_{0}/\hbar-\frac{\delta\varepsilon}{2}\sigma^{F}_{z}$ with the dominant part $H_{0}/\hbar=-\frac{\varepsilon}{2}\sigma^{F}_{z}-\frac{\Delta}{2}\sigma^{F}_{x}$~\cite{Science:Chiorescu2003,PRB:Yoshihara2014}. $(\sigma^{F}_{x},\sigma^{F}_{y},\sigma^{F}_{z})$ are the Pauli matrices for the flux qubit. $\varepsilon=\frac{2I_{p}\Phi_{0}}{\hbar}(\frac{\Phi_{ex}}{\Phi_{0}}-\frac{1}{2})$ is the frequency bias between $|L\rangle$ and $|R\rangle$ with the persistent current $I_{p}=0.3$ $\mu$A~\cite{PRL:Kakuyanagi2007} and the magnetic flux quantum $\Phi_{0}=\frac{\pi\hbar}{e}$. The interstate tunnel rate $\Delta$ depends on the specific physical realization of the flux qubit. To facilitate the future implementation of flux-qubit-ultracold-atom hybrids proposed in~\cite{PRA:Patton2013,RevMexFisS2011}, here we set $\Delta=2\pi\times6.8$ GHz which matches the clock transition of the $^{87}$Rb atom. $\delta\varepsilon$ denotes the qubit-energy fluctuations caused dominantly by the environmental flux noise. The effects of other noise sources (critical-current and charge noises) are not considered in this work~\cite{PRL:Kakuyanagi2007,PRL:Yoshihara2006}.

The eigenstates of $H_{0}/\hbar$ are derived as $|0\rangle=\cos\frac{\theta}{2}|L\rangle+\sin\frac{\theta}{2}|R\rangle$ and $|1\rangle=-\sin\frac{\theta}{2}|L\rangle+\cos\frac{\theta}{2}|R\rangle$, where $\theta$ is determined by $\cos\theta=\frac{\varepsilon}{E_{10}}$ and $\sin\theta=\frac{\Delta}{E_{10}}$ with the frequency spacing $E_{10}=\sqrt{\varepsilon^{2}+\Delta^{2}}$. In the basis of $|0\rangle$ and $|1\rangle$, we have $H/\hbar=\frac{E_{10}}{2}\Sigma^{F}_{z}+\frac{\delta\varepsilon}{2}(\cos\theta\Sigma^{F}_{z}+\sin\theta\Sigma^{F}_{x})$. The Heisenberg equations for the operators $\Sigma^{F}_{x}=\cos\theta\sigma^{F}_{x}-\sin\theta\sigma^{F}_{z}$, $\Sigma^{F}_{y}=-\sigma^{F}_{y}$, and $\Sigma^{F}_{z}=-\sin\theta\sigma^{F}_{x}-\cos\theta\sigma^{F}_{z}$ are given by
\begin{subequations}\label{FluxQubitEq}
\begin{eqnarray}
\dot{\Sigma}^{F}_{x}&=&-(E_{10}+\delta\varepsilon\cos\theta)\Sigma^{F}_{y},\\
\dot{\Sigma}^{F}_{y}&=&(E_{10}+\delta\varepsilon\cos\theta)\Sigma^{F}_{x}-\delta\varepsilon\sin\theta\Sigma^{F}_{z},\\
\dot{\Sigma}^{F}_{z}&=&\delta\varepsilon\sin\theta\Sigma^{F}_{y}.
\end{eqnarray}
\end{subequations}
According to the experimental measurements~\cite{PRL:Yoshihara2006,PRL:Kakuyanagi2007}, the noise spectrum density $S_{\delta\varepsilon}(f)=\int[\int\delta\varepsilon(t+\tau)\delta\varepsilon(t)dt]e^{-i2\pi f\tau}d\tau$ of $\delta\varepsilon$ exhibits two distinct regions: the $1/f$-type $S_{\delta\varepsilon}(f)=(\frac{2I_{p}\Phi_{0}}{\hbar})^{2}\frac{A_{\Phi}}{(2\pi f)^{0.9}}$ for the noise frequency $2\pi f\ll\Delta$ and the ohmic-dissipation $S_{\delta\varepsilon}(f)=\alpha(2\pi f)$ for $2\pi f\sim\Delta$.

The quantum behavior of the flux qubit can be simulated based on Eq.~(\ref{FluxQubitEq}), where $\delta\varepsilon$ is numerically generated based on the typical values of $A_{\Phi}=5\times10^{-12}$~\cite{PRL:Yoshihara2006} and $\alpha=10^{-5}$~\cite{RMP:Makhlin2001}. The energy-relaxation $T_{1}$ and dephasing $T_{2}$ times of the flux qubit are extracted by the decay of $|1\rangle$ and performing spin echo [Fig.~\ref{Fig1}(b)]. To verify the validity of the simulation method, we compare the numerical results with the analytical formulas derived from Fermi's golden rule, $T^{-1}_{1}=\pi\alpha E_{10}\sin^{2}\theta$~\cite{RMP:Makhlin2001,Book:2003} and $T^{-1}_{2}=\frac{1}{2}T^{-1}_{1}+(\frac{2I_{p}\Phi_{0}}{\hbar})\sqrt{A_{\Phi}\ln2}\cos\theta$~\cite{PRL:Yoshihara2006}. It is seen that they are well matched. $T_{2}$ strongly depends on the flux bias $\Phi_{ex}$ while $T_{1}$ is almost unchanged. At the optimal bias condition $\Phi_{ex}=2\Phi_{0}$ we obtain $T_{2}=2T_{1}$.

We focus on the flux-qubit Rabi oscillation between $|L\rangle$ and $|R\rangle$. Within the range of $\Phi_{ex}$ interested in this work [Fig.~\ref{Fig1}(c)], we have $(\frac{\varepsilon}{\Delta})^{2}\ll1$. For an ideal system with $\delta\varepsilon=0$, the persistent-current-flow direction switches alternately clockwise and counter-clockwise, i.e., $\langle\sigma^{F}_{z}(t)\rangle\simeq\cos E_{10}t$. This may be viewed as a local oscillator with the frequency of $\omega_{os}=E_{10}$. However, the nonzero $\delta\varepsilon$ disturbs $\omega_{os}$ around $E_{10}$, leading to the fluctuations $\delta\omega_{os}=\omega_{os}-E_{10}$, which gives rise to the limited $T_{1,2}$. We employ the Allan deviation $\sigma^{A}_{\omega_{os}}(\tau)$ to measure the stability of $\omega_{os}$ within an average time $\tau$~\cite{BOOK:Riehle2005}. As illustrated in Fig.~\ref{Fig1}(d), $\sigma^{A}_{\omega_{os}}(\tau)$ for different $\Phi_{ex}$ are nearly same for $\tau<1$ ns, corresponding to the fact that $T_{1}$ depends less on $\Phi_{ex}$. The $1/f$ component in $\delta\varepsilon$ mainly affects the relatively long-term ($\tau\gg1$ ns) stability of $\omega_{os}$, resulting in the strong decrease of $T_{2}$ for $\Phi_{ex}$ away from the optimal point. In the following, we set $\frac{\Phi_{ex}}{\Phi_{0}}-\frac{1}{2}=5\times10^{-4}$, where $T_{1}=1.5$ $\mu$s and $T_{2}=0.5$ $\mu$s. Stabilizing $\omega_{os}$ is equivalent to suppress $\delta\varepsilon$, which may potentially enhance $T_{1,2}$.

\begin{figure}[b]
\includegraphics[width=8.0cm]{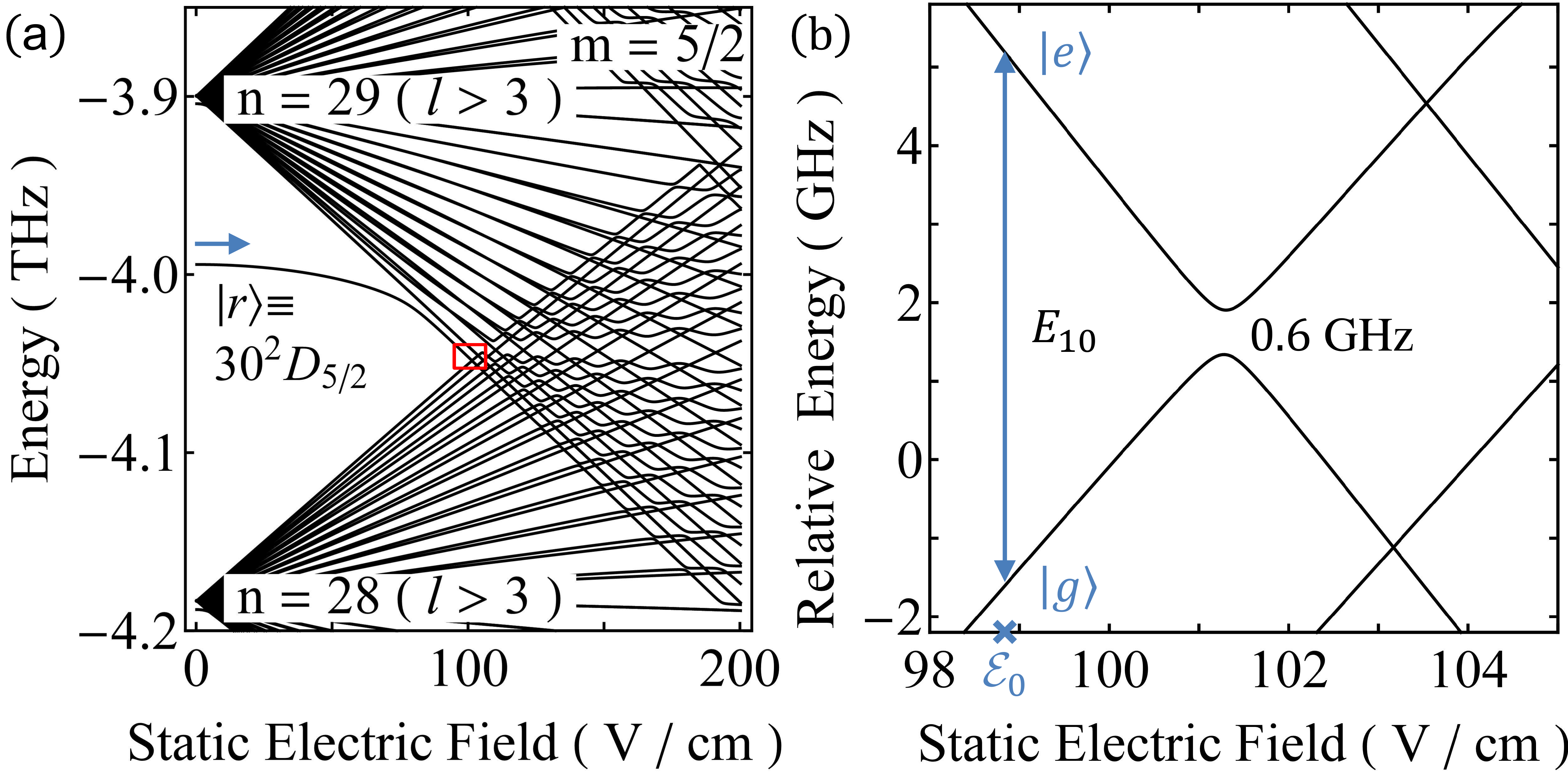}\\
\caption{(Color online) (a) dc Stark map of $^{87}$Rb around $|r\rangle\equiv30^{2}D_{5/2}(m=\frac{5}{2})$. The detail in the rectangle is displayed in (b), where the energy is relative to the value of 4.05 THz. At ${\cal{E}}_{0}=98.8$ V/cm, the frequency spacing between adiabatic $|e\rangle$ and $|g\rangle$ eigenstates is equal to $E_{10}$.}\label{Fig2}
\end{figure}

\textit{Frequency discrimination.} We utilize the flux-qubit-resonator-atom hybrid platform established in~\cite{QST:Yu2017} to discriminate against $\delta\omega_{os}$ of $\omega_{os}$. As depicted in Fig.~\ref{Fig1}(a), a series LC resonator is biased by a constant voltage source and weakly coupled to the flux qubit with a mutual inductance of $M=3$ pH~\cite{Nature:Chiorescu2004}. The resonator's capacitor $C=0.1$ pF comprises of a large $C_{l}$ and a small $C_{s}$ connected in parallel, where $C_{s}=4$ fF is formed by a pair of identical parallel plates with the area of $\Delta x\times\Delta y=0.2\times0.2$ mm$^{2}$ and interplate distance of $\Delta z=0.1$ mm. The resonator frequency $\omega_{LC}=\frac{1}{\sqrt{LC}}$ is set nearly resonant to $\omega_{os}$ with a detuning $\delta_{f}=\omega_{LC}-\omega_{os}=\delta_{f,0}+\delta\omega_{os}$ and $\delta_{f,0}=\omega_{LC}-E_{10}$, resulting in the inductance $L\approx5$ nH~\cite{APL:Vissers2015}. We assume that the constant voltage source does not introduce extra noises.

An ensemble of $^{87}$Rb Rydberg atoms (number of $n_{at}$) fly through $C_{s}$ in the $x$-direction at the same velocity $v$. Within $C_{s}$, the atoms interact with the nearly homogeneously-distributed intracapacitor electric field. This $z$-direction field contains two components: the static ${\cal{E}}_{0}$, which is produced by the constant voltage bias and tunes the energy spectrum of the atom, and the oscillating ${\cal{E}}$, whose amplitude depends on the detuning $\delta_{f}$. Figure~\ref{Fig2}(a) displays the dc Stark map of the Rydberg atom around $|r\rangle\equiv30^{2}D_{5/2}(m=\frac{5}{2})$. We focus on two adiabatic energy curves which start respectively from $|r\rangle$ and a manifold state composed of a set of $|n=20,l\leq3,j=l\pm\frac{1}{2},m=\frac{5}{2}\rangle$ states at the zero field. An energy-level avoided crossing (frequency gap $2\pi\times0.6$ GHz) occurs between two curves at the static electric field of 101.3 V/cm [Fig.~\ref{Fig2}(b)].

We set ${\cal{E}}_{0}=98.8$ V/cm, where the frequency spacing between two corresponding adiabatic eigenstates (labeled as $|e\rangle$ and $|g\rangle$) is equal to $E_{10}$. Since other adiabatic states are far apart from $|e\rangle$ and $|g\rangle$, the Rydberg atom may be viewed as a two-level system. The radius of Rydberg atom is 65 nm~\cite{JPB:Low2012} and the zero-Kelvin lifetime is 25 $\mu$s~\cite{JPB:Branden2012}. The large atom-surface separation (50 $\mu$m) ensures that the stray fields from the SC chip hardly influence the Rydberg states~\cite{PRA:Crosse2010}. For $n_{at}=200$, the average separation among atoms is 15 $\mu$m, large enough to suppress the interatomic interactions.

All atoms, before entering $C_{s}$, are prepared in $|r\rangle$ via the two-photon $5^{2}S_{1/2}(m=\frac{1}{2})-5^{2}P_{3/2}(m=\frac{3}{2})-|r\rangle$ excitation by using 780 nm and 480 nm laser lights [Fig.~\ref{Fig1}(a)]. When approaching $C_{s}$, the atoms adiabatically follow the corresponding energy curve due to the static fringe field~\cite{PRA:Yu2017}. The oscillating ${\cal{E}}$ field couples to the $|e\rangle-|g\rangle$ transition for a time duration of $t_{int}=\Delta x/v$. The atomic transition rate relies on the flux-qubit-resonator detuning $\delta_{f}$. As the atoms fly away from $C_{s}$, the component in $|e\rangle$ adiabatically evolves back to $|r\rangle$. Then, the atoms in $|r\rangle$ are completely mapped into $5^{2}S_{1/2}(m=\frac{1}{2})$ at a rapid rate ($\sim$ns~\cite{PRL:Huber2011}) via the two-photon transition again. The resulting atomic population is measured by the fluorescence method based on the $5^{2}S_{1/2}(m=\frac{1}{2})-5^{2}P_{3/2}(m=\frac{3}{2})$ transition whose decay rate is $\gamma=2\pi\times6.1$ MHz. From the number $n_{ph}$ of fluorescence photons collected by a photodetector during a time length of $t_{d}$, the fluctuations $\delta\omega_{os}$ may be derived. The measured result is fed back into the current source via the servo to stabilize the flux bias [Fig.~\ref{Fig1}(a)].

Next, we derive the frequency-discrimination curve (FDC). The Hamiltonian describing the interface between atoms and ${\cal{E}}$ is expressed as $\tilde{H}/\hbar=\frac{\omega_{eg}}{2}\sigma^{A}_{z}+\Omega\sigma^{A}_{x}$, where $\sigma^{A}_{x}=|e\rangle\langle g|+|g\rangle\langle e|$ and $\sigma^{A}_{z}=|e\rangle\langle e|-|g\rangle\langle g|$ are the Pauli matrices for the atoms and the Rabi frequency is defined as $\Omega=-d_{0}{\cal{E}}/\hbar$ with $d_{0}\sim700$ $ea_{0}$~\cite{PRA:Yu2016-2,Book:Gallagher}. The equations of motion of atomic density matrix elements $\rho_{\mu\nu}=\langle\mu|\rho|\nu\rangle$  with $\mu,\nu=e,g$ are derived as
\begin{subequations}\label{AtomEq}
\begin{eqnarray}
\dot{\rho}_{ee}&=&i\Omega^{\ast}\rho_{eg}-i\Omega\rho^{\ast}_{eg},\\
\dot{\rho}_{eg}&=&-iE_{10}\rho_{eg}+i\Omega^{\ast}(2\rho_{ee}-1).
\end{eqnarray}
\end{subequations}
$\rho_{ee}$ denotes the probability of the atoms being in $|e\rangle$ and $\rho_{eg}$ corresponds to the atomic polarizability.

The Kirchoff’s Laws lead to the wave equation for ${\cal{E}}$
\begin{equation}\label{FieldEq}
\textstyle{\ddot{{\cal{E}}}+\kappa\dot{{\cal{E}}}+\omega^{2}_{LC}{\cal{E}}=-\frac{C_{s}}{C}\frac{\ddot{{\cal{P}}}}{\varepsilon_{0}}+\omega^{2}_{LC}\frac{MI_{p}}{\Delta z}\dot{\sigma}^{F}_{z}}.
\end{equation}
where ${\cal{P}}=\frac{n_{at}d_{0}}{V_{\textrm{eff}}}(\rho_{eg}+\rho^{\ast}_{eg})$ is the polarization density of the atoms and $V_{\textrm{eff}}$ is the effective resonator-mode volume. In the limits of $C\gg C_{s}$ and $V_{\textrm{eff}}\gg\Delta x\Delta y\Delta z$, the atoms hardly affect ${\cal{E}}$. The second term on the right side of Eq.~(\ref{FieldEq}), originating from the flux-qubit-resonator coupling, plays a role of current driving source. To weaken the resonator filtering effect, the resonator loss rate is chosen to be $\kappa=4\gamma$, which corresponds to a low $Q$-factor of 300. Moreover, the frequency bias $\varepsilon$ should be rewritten as $\varepsilon=\frac{2I_{p}\Phi_{0}}{\hbar}(\frac{\Phi_{ex}}{\Phi_{0}}-\frac{1}{2}+\frac{MI}{\Phi_{0}})$, where the current $I$ flowing in the resonator is given by $I=\Delta zC\frac{d}{dt}({\cal{E}}+\frac{C_{s}}{C}\frac{{\cal{P}}}{\varepsilon_{0}})$.

\begin{figure}[b]
\includegraphics[width=8.0cm]{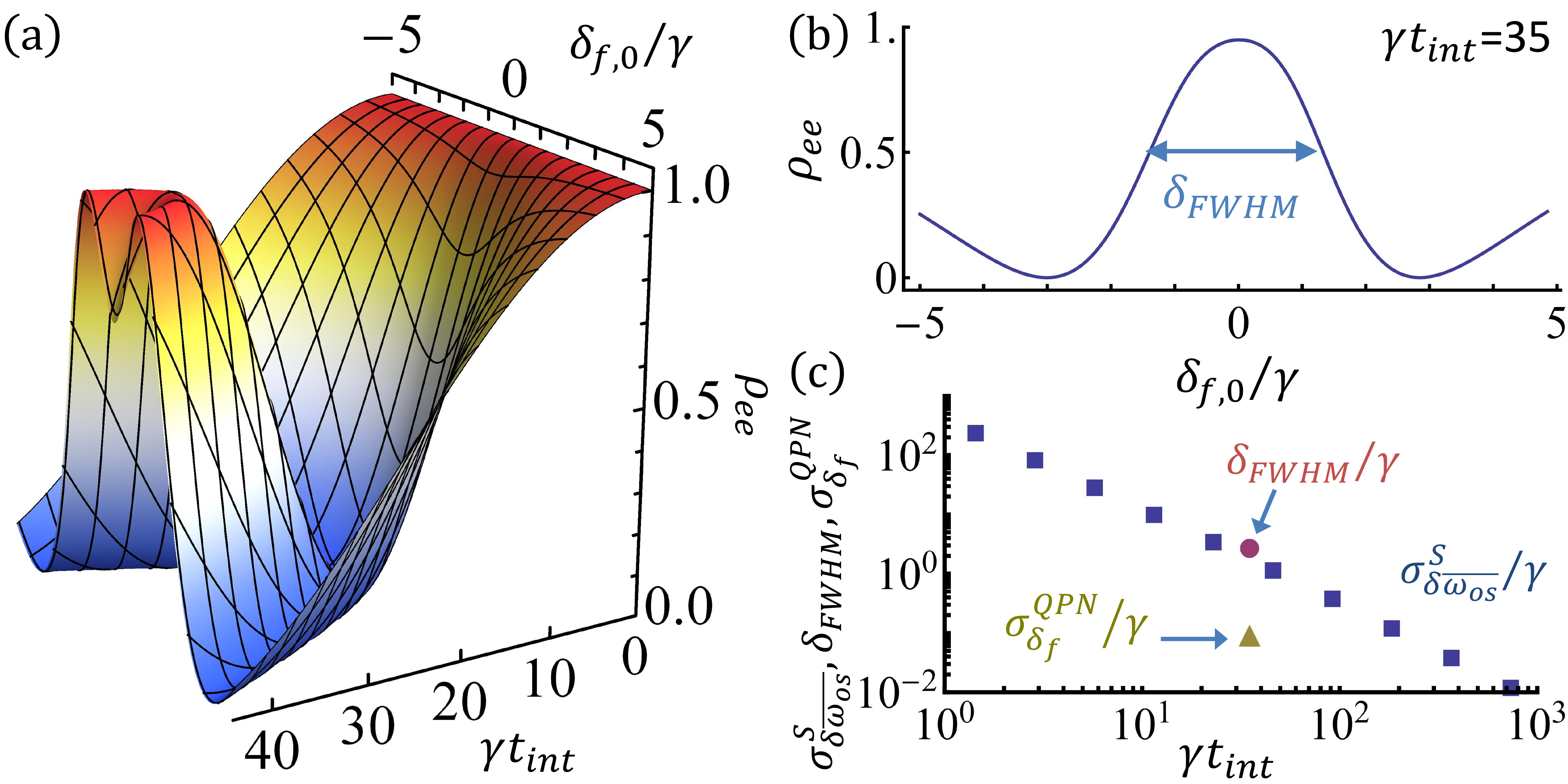}\\
\caption{(Color online) (a) Probability $\rho_{ee}$ as a function of the flux qubit-resonator detuning $\delta_{f,0}$ and the atom-resonator interaction time $t_{int}$. (b) Frequency-discrimination curve, $\rho_{ee}$ vs. $\delta_{f,0}$, at $\gamma t_{int}=35$. (c) Standard deviation $\sigma^{S}_{\delta\overline{\omega_{os}}}$ of the average fluctuations $\delta\overline{\omega_{os}}$ within $t_{int}$. The projection-noise-induced uncertainty $\sigma^{QPN}_{\delta_{f}}$ and the full width at half maximum $\delta_{FWHM}$ of the frequency-discrimination curve of (b) are also inserted in (c).}\label{Fig3}
\end{figure}

Combing Eqs.~(\ref{FluxQubitEq})-(\ref{FieldEq}), one can simulate the dynamics of the whole hybrid system. Figure~\ref{Fig3}(a) depicts the dependence of the probability $\rho_{ee}$ on the detuning $\delta_{f,0}$ and the interaction time $t_{int}$, where $\delta\omega_{os}=0$. It is seen that as $t_{int}$ is increased, $\rho_{ee}$ exhibits the oscillating behavior, i.e, the Rabi oscillation of atoms. The effect of $\delta_{f,0}$ also merges evidently, i.e., $\rho_{ee}$ for the resonant flux-qubit-resonator coupling ($\delta_{f,0}=0$) varies much faster than the non-resonant ($\delta_{f,0}\neq0$) case, for a long enough $t_{int}$. Indeedly, this dependence is caused by the frequency discrimination characteristics of the resonator.

For a certain $t_{int}$, one obtains the one-to-one correspondence between $\rho_{ee}$ and $\delta_{f,0}$ on the negative side of $\delta_{f,0}$, which may be employed to discriminate against the average fluctuation $\delta\overline{\omega_{os}}=\frac{1}{t_{int}}\int^{t_{int}}_{0}\delta\omega_{os}(t)dt$ of $\omega_{os}$ within $t_{int}$. The optimal $t_{int}$ relies on the requisite that the standard deviation $\sigma^{S}_{\delta\overline{\omega_{os}}}$ of $\delta\overline{\omega_{os}}$ must satisfy $\sigma^{S}_{\delta\overline{\omega_{os}}}<\delta_{FWHM}$, where $\delta_{FWHM}$ is the full width at half maximum of the FDC. Otherwise, $\delta\overline{\omega_{os}}$ will not be uniquely identified. $\sigma^{S}_{\delta\overline{\omega_{os}}}$ should also be larger than the uncertainty $\sigma^{QPN}_{\delta_{f}}=\frac{\delta_{FWHM}}{2\sqrt{n_{at}}}$ induced by the quantum projection noise (QPN)~\cite{PRA:Itano1993} occurring in measuring the atomic population in $|e\rangle$. We set $\gamma t_{int}=35$ with $v=180$ m/s. The corresponding FDC is plotted in Fig.~\ref{Fig3}(b), where the stabilization point is commonly chosen at the highest-gradient position, $\delta_{f,0}=-\frac{\delta_{FWHM}}{2}$. Figure~\ref{Fig3}(c) shows the comparison among $\delta\overline{\omega_{os}}$, $\delta_{FWHM}$, and $\sigma^{QPN}_{\delta_{f}}$. 

\begin{figure}
\includegraphics[width=8.0cm]{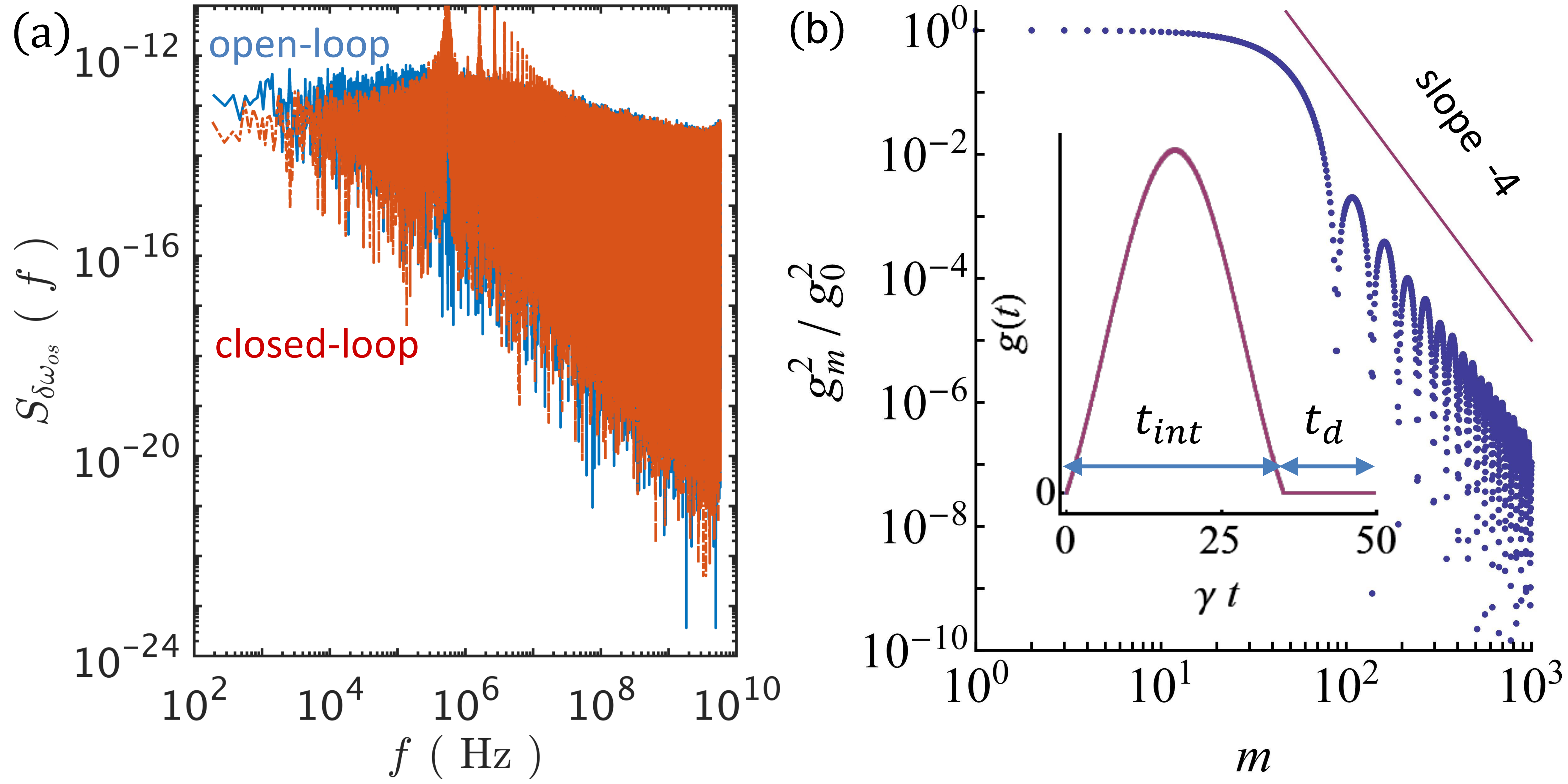}\\
\caption{(Color online) (a) Power spectral density $S_{\delta\omega_{os}}(f)$ of the fluctuations $\delta\omega_{os}$ of open- and closed-loop systems. (b) Spectrum of $g(t)/g_{0}$. Inset: Sensitivity function $g(t)$ within a feedback cycle.}\label{Fig4}
\end{figure}

\textit{Clock operation.} Based on the obtained FDC, one can derive $\delta\overline{\omega_{os}}$ from the atomic-population measurement which is disturbed by the unavoidable QPN. Another fundamental noise occurring in the population measurement is the photon shot noise (SN) arising from the particle-like nature of light~\cite{MPLB:Beenakker1999}. The corresponding signal-to-noise ratio is given by $\frac{1}{\sqrt{n_{ph}}}$~\cite{APB:Bjorklund1983}. To suppress the SN, we set $\gamma t_{d}=15$, resulting in $n_{ph}=15$ for an efficient collection of fluorescence photons.

From the derived $\delta\overline{\omega_{os}}$, one can further calculate the average value $\delta\overline{\varepsilon}$ of $\delta\varepsilon$ within $t_{int}$, i.e., $\delta\overline{\varepsilon}=\sqrt{(E_{10}+\delta\overline{\omega_{os}})^{2}-\Delta^{2}}-\varepsilon$. $\delta\overline{\varepsilon}$ may be compensated by tuning $\varepsilon$ via feedback controlling $\Phi_{ex}$ at the end of photon collection. Repeating the whole process leads to a stabilized persistent-current Rabi oscillation. The feedback-control cycle $T_{c}=1.3$ $\mu$s consists of an atom-resonator-interaction duration $t_{int}$ and a dead-time period $t_{d}$ for collecting fluorescence photons. Suppressing $T_{c}$ requires a larger $d_{0}$, i.e., higher Rydberg states, and a faster decay rate $\gamma$. It should be noted that since $t_{d}$ occupies over one third of $T_{c}$, the feedback-control efficiency is reduced.

Combing Eqs. (\ref{FluxQubitEq})-(\ref{FieldEq}), we numerically perform the clock running, where $\varepsilon$ in Eq.~(\ref{FluxQubitEq}) is corrected by $\delta\overline{\varepsilon}$ every $T_{c}$. In Fig.~\ref{Fig4}(a), we compare the spectral density $S_{\delta\omega_{os}}(f)=\int[\int\delta\omega_{os}(t+\tau)\delta\omega_{os}(t)dt]e^{-i2\pi f\tau}d\tau$ of $\delta\omega_{os}$ for both open- and closed-loop systems. It is seen that the feedback control barely affects $S_{\delta\omega_{os}}(f)$ for $f>T^{-1}_{c}\sim1$ MHz while suppresses the fluctuations in the closed-loop system in the low-$f$ ($f<T^{-1}_{c}$) regime. The corresponding $\sigma^{A}_{\omega_{os}}(\tau)$ of the clock operation is displayed in Fig.~\ref{Fig1}(d). As one can see, the closed-loop $\sigma^{A}_{\omega_{os}}(\tau)$ is lower than that of open-loop system after $\tau>T_{c}$, indicating the improvement of frequency stability of persistent-current oscillator. However, we should note that the energy-relaxation time $T_{1}$ of the closed-loop flux qubit is not extended since $T_{c}$ is similar to the free-running $T_{1}$.

According to~\cite{PRL:Santarelli1999}, the stability of $\omega_{os}$ is limited, in principle, by $\sigma^{A}_{\omega_{os},L}(\tau)=\frac{1}{Q}\sqrt{\frac{T_{c}}{\tau}}(\frac{1}{n_{at}}+\frac{1}{n_{at}n_{ph}})^{1/2}=\frac{2.1\times10^{-7}}{\sqrt{\tau}}$, where $Q=\frac{E_{10}}{\delta_{FWHM}}=400$ denotes the $Q$-factor of the FDC, and the first term in the brackets is the QPN while the second term corresponds to the SN occurring in the fluorescence detection. However, as demonstrated in Fig.~\ref{Fig1}(d), $\sigma^{A}_{\omega_{os},L}(\tau)$ is much lower than the closed-loop $\sigma^{A}_{\omega_{os}}(\tau)$. This is because the Dick effect, induced by the interrogation frequency noise and nonzero dead time in the feedback cycle, strongly degrades the oscillator's stability~\cite{NatPhoton:Takamoto2011}. Following~\cite{IEEE:Santarelli1998}, the Dick-effect-limited Allan deviation is expressed as $\sigma^{A}_{\omega_{os},D}(\tau)=[\frac{1}{\tau}\sum_{m=1}^{\infty}\frac{g^{2}_{m}}{g^{2}_{0}}S_{\delta\omega_{os}}(\frac{m}{T_{c}})]^{1/2}=\frac{7.9\times10^{-7}}{\sqrt{\tau}}$, where $g_{0}=\tfrac{1}{T_{c}}\int_{0}^{T_{c}}g(t)dt$, $g^{2}_{m}=g^{2}_{c,m}+g^{2}_{s,m}$, $g_{c,m}=\tfrac{1}{T_{c}}\int_{0}^{T_{c}}g(t)\cos\tfrac{2\pi mt}{T_{c}}dt$, and $g_{s,m}=\tfrac{1}{T_{c}}\int_{0}^{T_{c}}g(t)\sin\tfrac{2\pi mt}{T_{c}}dt$. The sensitivity function $g(t)$ to the frequency fluctuations of the flux-qubit Rabi oscillation, which may be assumed to have the form $I_{p}\cos E_{10}t$, is defined as $g(t)=2\lim_{\Delta\phi\rightarrow0}\frac{\delta\rho_{ee}(t)}{\Delta\phi}$. $\delta\rho_{ee}(t)$ denotes the change of the probability of the atoms being in $|e\rangle$, caused by an infinitesimally small phase step $\Delta\phi$ of the oscillator signal, i.e., $\cos E_{10}t\rightarrow\cos(E_{10}t+\Delta\phi)$, arising at time $t$.

Figure~\ref{Fig4}(b) depicts the numerical result of $g(t)$, which is similar to that of the common Rabi interrogation in optical lattice clocks~\cite{PRA:Al-Masoudi2015}, and $(g^{2}_{m}/g^{2}_{0})$ vs. $m$. It is seen that the low-frequency fluctuations of $\omega_{os}$ primarily affects $\sigma^{A}_{\omega_{os},D}(\tau)$. The resulting $\sigma^{A}_{\omega_{os},D}(\tau)$ is inserted in Fig.~\ref{Fig1}(d). which proves that the long-term stability of $\omega_{os}$ is limited by the Dick effect. Reducing the dead time $t_{d}$ may suppress $\sigma^{A}_{\omega_{os},D}(\tau)$. However, the shorter $t_{d}$ leads to the decrease of $n_{ph}$, which raises $\sigma^{A}_{\omega_{os},L}(\tau)$.

\textit{Conclusion.} We have investigated the stabilization of a flux qubit by the clock-stability-transfer technique, where the periodic persistent-current oscillation is locked to a microwave Rydberg-Rydberg transition of the $^{87}$Rb atom via the resonator serving as a quantum bus. The SC-qubit-resonator-atom hybrid platform may be generalized to other type of SC qubits. So far, the coherent SC-qubit-resonator coupling has been well demonstrated in experiments~\cite{Nature:Chiorescu2004,Science:Wang2016,NatCommun:Yan2016}. In contrast, the implementation of SC-circuit-atom interface is very limited, despite of plenty of relevant theoretical proposals~\cite{RMP:Xiang2013,SciRep:Yu2016,QST:Yu2017,ProcSPIE:Hufnagel2017,PRL:Petrosyan2008,PRA:Patton2013,PRA:Yu2016-1,PRA:Yu2016-2}. The main challenge is manipulating neutral atoms nearby the cryogenic surface. Nevertheless, the recent experiments~\cite{PRA:Thiele2014,PRA:Stammeier2017,NatCommun:Hattermann2017} show that coupling ultracold (Rydberg) atoms to a SC resonator is attainable. The fast development of low-temperature electronics further combined with the clock technology may potentially be widely applied in SC quantum information processing, allowing it to be immune to environmental noises~\cite{Nature:Vijay2012}.

\begin{acknowledgments}
D. Y. would like to thank Hidetoshi Katori for the support of the feedback-control code. This research has been supported by the National Research Foundation Singapore \& by the Ministry of Education Singapore Academic Research Fund Tier 2 (Grant No. MOE2015-T2-1-101).
\end{acknowledgments}


\begin{thebibliography}{10}

\bibitem{RMP:Xiang2013} Z.-L. Xiang, S. Ashhab, J. Q. You, and F. Nori, Rev. Mod. Phys. {\bf 85}, 623 (2013). \url{https://link.aps.org/doi/10.1103/RevModPhys.85.623}

\bibitem{SciRep:Yu2016} D. Yu, M. M. Valado, C. Hufnagel, L. C. Kwek, L. Amico, and R. Dumke, Sci. Rep. {\bf 6}, 38356 (2016). \url{http://dx.doi.org/10.1038/srep38356}

\bibitem{QST:Yu2017} D. Yu, L. C. Kwek, L. Amico, and R. Dumke, Quantum Sci. Technol. {\bf 2}, 035005 (2017). \url{http://stacks.iop.org/2058-9565/2/i=3/a=035005}

\bibitem{ProcSPIE:Hufnagel2017} C. Hufnagel, A. Landra, L. C. Chean, D. Yu, and R. Dumke, Proc. SPIE {\bf 10358}, 10358 (2017). \url{https://doi.org/10.1117/12.2275929}

\bibitem{PRL:Petrosyan2008} D. Petrosyan and M. Fleischhauer, Phys. Rev. Lett. {\bf 100}, 170501 (2008). \url{https://link.aps.org/doi/10.1103/PhysRevLett.100.170501}

\bibitem{PRA:Patton2013} K. R. Patton and U. R. Fischer, Phys. Rev. A {\bf 87}, 052303 (2013). \url{https://link.aps.org/doi/10.1103/PhysRevA.87.052303}

\bibitem{PRA:Yu2016-1} D. Yu, M. M. Valado, C. Hufnagel, L. C. Kwek, L. Amico, and R. Dumke, Phys. Rev. A {\bf 93}, 042329 (2016). \url{https://link.aps.org/doi/10.1103/PhysRevA.93.042329}

\bibitem{PRA:Yu2016-2} D. Yu, A. Landra, M. M. Valado, C. Hufnagel, L. C. Kwek, L. Amico, and R. Dumke, Phys. Rev. A {\bf 94}, 062301 (2016). \url{https://link.aps.org/doi/10.1103/PhysRevA.94.062301}

\bibitem{NatPhys:Niemczyk2010} T. Niemczyk, F. Deppe, H. Huebl, E. P. Menzel, F. Hocke, M. J. Schwarz, J. J. Garcia-Ripoll, D. Zueco, T. H\"{u}mmer, E. Solano, A. Marx, and R. Gross, Nat. Phys. {\bf 6}, 772 (2010). \url{http://dx.doi.org/10.1038/nphys1730}

\bibitem{NatPhys:Yoshihara2017} F. Yoshihara, T. Fuse, S. Ashhab, K. Kakuyanagi, S. Saito, and K. Semba, Nat. Phys. {\bf 13}, 44 (2017). \url{http://dx.doi.org/10.1038/nphys3906}

\bibitem{PRA:Yu2017} D. Yu, L. C. Kwek, L. Amico, and R. Dumke, Phys. Rev. A {\bf 95}, 053811 (2017). \url{https://link.aps.org/doi/10.1103/PhysRevA.95.053811}

\bibitem{RevMexFisS2011} J. E. Hoffman, J. A. Grover, Z. Kim, A. K. Wood, J. R. Anderson, A. J. Dragt, M. Hafezi, C. J. Lobb, L. A. Orozco, S. L. Rolston, J. M. Taylor, C. P. Vlahacos, and F. C. Wellstood, Rev. Mex. F\'is. S {\bf 57}, 1 (2011); also at arXiv:1108.4153 [quant-ph]. \url{https://rmf.smf.mx/pdf/rmf-s/57/3/57_3_1.pdf}

\bibitem{NatCommun:Bernon2013} S. Bernon, H. Hattermann, D. Bothner, M. Knufinke, P. Weiss, F. Jessen, D. Cano, M. Kemmler, R. Kleiner, D. Koelle, and J. Fort\'{a}gh, Nat. Commun. {\bf 4}, 2380 (2013). \url{http://dx.doi.org/10.1038/ncomms3380}

\bibitem{PRL:Weiss2015} P. Weiss, M. Knufinke, S. Bernon, D. Bothner, L. S\'{a}rk\'{a}ny, C. Zimmermann, R. Kleiner, D. Koelle, J. Fort\'{a}gh, and H. Hattermann, Phys. Rev. Lett. {\bf 114}, 113003 (2015). \url{https://link.aps.org/doi/10.1103/PhysRevLett.114.113003}

\bibitem{Nature:Takamoto2005} M. Takamoto, F.-L. Hong, R. Higashi, and H. Katori, Nature (London) {\bf 435}, 321 (2005). \url{http://dx.doi.org/10.1038/nature03541}

\bibitem{NatPhys:Akatsuka2008} T. Akatsuka, M. Takamoto, and H. Katori, Nat. Phys. {\bf 4}, 954 (2008). \url{http://dx.doi.org/10.1038/nphys1108}

\bibitem{NatPhoton:Ushijima2015} I. Ushijima, M. Takamoto, M. Das, T. Ohkubo, and H. Katori, Nat. Photon. {\bf 9}, 185 (2015). \url{http://dx.doi.org/10.1038/nphoton.2015.5}

\bibitem{PRL:Astafiev2004} O. Astafiev, Yu. A. Pashkin, Y. Nakamura, T. Yamamoto, and J. S. Tsai, Phys. Rev. Lett. {\bf 93}, 267007 (2004). \url{https://link.aps.org/doi/10.1103/PhysRevLett.93.267007}

\bibitem{PRB:Ithier2005} G. Ithier, E. Collin, P. Joyez, P. J. Meeson, D. Vion, D. Esteve, F. Chiarello, A. Shnirman, Y. Makhlin, J. Schriefl, and G. Sch\"{o}n, Phys. Rev. B {\bf 72}, 134519 (2005). \url{https://link.aps.org/doi/10.1103/PhysRevB.72.134519}

\bibitem{PRB:Lupascu2005} A. Lupa\ifmmode \mbox{\c{s}}\else \c{s}\fi{}cu, C. J. P. M. Harmans, and J. E. Mooij, Phys. Rev. B {\bf 71}, 184506 (2005). \url{https://link.aps.org/doi/10.1103/PhysRevB.71.184506}

\bibitem{PRL:Anton2013} S. M. Anton, J. S. Birenbaum, S. R. O'Kelley, V. Bolkhovsky, D. A. Braje, G. Fitch, M. Neeley, G. C. Hilton, H.-M. Cho, K. D. Irwin, F. C. Wellstood, W. D. Oliver, A. Shnirman, and J. Clarke, Phys. Rev. Lett. {\bf 110}, 147002 (2013). \url{https://link.aps.org/doi/10.1103/PhysRevLett.110.147002}

\bibitem{NJP:Yu2017} D. Yu, A. Landra, L. C. Kwek, L. Amico, and R. Dumke, New J. Phys. {\bf 20}, 023031 (2018). \url{http://stacks.iop.org/1367-2630/20/i=2/a=023031}

\bibitem{Science:Mooij1999} J. E. Mooij, T. P. Orlando, L. Levitov, L. Tian, C. H. van der Wal, and S. Lloyd, Science {\bf 285}, 1036 (1999). \url{http://science.sciencemag.org/content/285/5430/1036}

\bibitem{Science:Chiorescu2003} I. Chiorescu, Y. Nakamura, C. J. P. M. Harmans, and J. E. Mooij, Science {\bf 299}, 1869 (2003). \url{http://science.sciencemag.org/content/299/5614/1869}

\bibitem{PRB:Yoshihara2014} F. Yoshihara, Y. Nakamura, F. Yan, S. Gustavsson, J. Bylander, W. D. Oliver, and J.-S. Tsai, Phys. Rev. B {\bf 89}, 020503 (2014). \url{https://link.aps.org/doi/10.1103/PhysRevB.89.020503}

\bibitem{PRL:Kakuyanagi2007} K. Kakuyanagi, T. Meno, S. Saito, H. Nakano, K. Semba, H. Takayanagi, F. Deppe, and A. Shnirman, Phys. Rev. Lett. {\bf 98}, 047004 (2007). \url{https://link.aps.org/doi/10.1103/PhysRevLett.98.047004}

\bibitem{PRL:Yoshihara2006} F. Yoshihara, K. Harrabi, A. O. Niskanen, Y. Nakamura, and J. S. Tsai, Phys. Rev. Lett. {\bf 97}, 167001 (2006). \url{https://link.aps.org/doi/10.1103/PhysRevLett.97.167001}

\bibitem{RMP:Makhlin2001} Y. Makhlin, G. Sch\"{o}n, and A. Shnirman, Rev. Mod. Phys. {\bf 73}, 357 (2001). \url{https://link.aps.org/doi/10.1103/RevModPhys.73.357}

\bibitem{Book:2003} Y. Makhlin, G. Sch\"{o}n, and A. Shnirman, in {\it New Directions in Mesoscopic Physics (Towards Nanoscience)}, edited by R. Fazio, V. F. Gantmakher, and Y. Imry (Kluwer, Dordrecht, 2003), pp. 197--224.

\bibitem{BOOK:Riehle2005} F. Riehle, \textit{Frequency Standards: Basics and Applications} (New York: Wiley, 2004).

\bibitem{Nature:Chiorescu2004} I. Chiorescu, P. Bertet, K. Semba, Y. Nakamura, C. J. P. M. Harmans, and J. E. Mooij, Nature (London) {\bf 431}, 159 (2004). \url{http://dx.doi.org/10.1038/nature02831}

\bibitem{APL:Vissers2015} M. R. Vissers, J. Hubmayr, M. Sandberg, S. Chaudhuri, C. Bockstiegel, and J. Gao, Appl. Phys. Lett. {\bf 107}, 062601 (2015). \url{https://doi.org/10.1063/1.4927444}

\bibitem{JPB:Low2012} R. L\"{o}w, H. Weimer, J. Nipper, J. B. Balewski, B. Butscher, H. P. B\"{u}chler, and T. Pfau, J. Phys. B: At. Mol. Opt. Phys. {\bf 45}, 113001 (2012). \url{http://stacks.iop.org/0953-4075/45/i=11/a=113001}

\bibitem{JPB:Branden2012} D. B. Branden, T. Juhasz, T. Mahlokozera, C. Vesa, R. O. Wilson, M. Zheng, A. Kortyna, and D. A. Tate, J. Phys. B: At. Mol. Opt. Phys. {\bf 43}, 015002 (2012). \url{http://stacks.iop.org/0953-4075/43/i=1/a=015002}

\bibitem{PRA:Crosse2010} J. A. Crosse, S. \AA{}. Ellingsen, K. Clements, S. Y. Buhmann, and S. Scheel, Phys. Rev. A {\bf 82}, 010901 (2010). \url{https://link.aps.org/doi/10.1103/PhysRevA.82.010901}

\bibitem{PRL:Huber2011} B. Huber, T. Baluktsian, M. Schlagm\"{u}ller, A. K\"{o}lle, H. K\"{u}bler, R. L\"{o}w, and T. Pfau, Phys. Rev. Lett. {\bf 107}, 243001 (2011). \url{https://link.aps.org/doi/10.1103/PhysRevLett.107.243001}

\bibitem{Book:Gallagher} T. F. Gallagher, {\it Rydberg Atoms} (New York: Cambridge University Press, 1994).

\bibitem{PRA:Itano1993} W. M. Itano, J. C. Bergquist, J. J. Bollinger, J. M. Gilligan, D. J. Heinzen, F. L. Moore, M. G. Raizen, and D. J. Wineland, Phys. Rev. A {\bf 47}, 3554 (1993). \url{https://link.aps.org/doi/10.1103/PhysRevA.47.3554}

\bibitem{MPLB:Beenakker1999} C. W. J. Beenakker and M. Patra, Mod. Phys. Lett. B {\bf 13}, 337 (1999). \url{https://doi.org/10.1142/S0217984999000439}

\bibitem{APB:Bjorklund1983} G. C. Bjorklund, M. D. Levenson, W. Lenth, and C. Ortiz, Appl. Phys. B {\bf 32}, 145 (1983). \url{https://doi.org/10.1007/BF00688820}

\bibitem{PRL:Santarelli1999} G. Santarelli, Ph. Laurent, P. Lemonde, A. Clairon, A. G. Mann, S. Chang, A. N. Luiten, and C. Salomon, Phys. Rev. Lett. {\bf 82}, 4619 (1999). \url{https://link.aps.org/doi/10.1103/PhysRevLett.82.4619}

\bibitem{NatPhoton:Takamoto2011} M. Takamoto, T. Takano, and H. Katori, Nat. Photon. {\bf 5}, 288 (2011). \url{http://dx.doi.org/10.1038/nphoton.2011.34}

\bibitem{IEEE:Santarelli1998} G. Santarelli, C. Audoin, A. Makdissi, P. Laurent, G. J. Dick, and A. Clairon, IEEE Trans. Ultrason. Ferroelectr. Freq. Control {\bf 45}, 887 (1998). \url{https://doi.org/10.1109/58.710548}

\bibitem{PRA:Al-Masoudi2015} A. Al-Masoudi, S. D\"{o}rscher, S. H\"{a}fner, U. Sterr, and C. Lisdat, Phys. Rev. A {\bf 92}, 063814 (2015). \url{https://link.aps.org/doi/10.1103/PhysRevA.92.063814}

\bibitem{Science:Wang2016} C. Wang, Y. Y. Gao, P. Reinhold, R. W. Heeres, N. Ofek, K. Chou, C. Axline, M. Reagor, J. Blumoff, K. M. Sliwa, L. Frunzio, S. M. Girvin, L. Jiang, M. Mirrahimi, M. H. Devoret, R. J. Schoelkopf, Science {\bf 352}, 1087 (2016). \url{http://science.sciencemag.org/content/352/6289/1087}

\bibitem{NatCommun:Yan2016} F. Yan, S. Gustavsson, A. Kamal, J. Birenbaum, A. P. Sears, D. Hover, T. J. Gudmundsen, D. Rosenberg, G. Samach, S. Weber, J. L. Yoder, T. P. Orlando, J. Clarke, A. J. Kerman, and W. D. Oliver, Nat. Commun. {\bf 7}, 12964 (2016). \url{http://dx.doi.org/10.1038/ncomms12964}

\bibitem{PRA:Thiele2014} T. Thiele, S. Filipp, J. A. Agner, H. Schmutz, J. Deiglmayr, M. Stammeier, P. Allmendinger, F. Merkt, and A. Wallraff, Phys. Rev. A {\bf 90}, 013414 (2014). \url{https://link.aps.org/doi/10.1103/PhysRevA.90.013414}

\bibitem{PRA:Stammeier2017} M. Stammeier, S. Garcia, T. Thiele, J. Deiglmayr, J. A. Agner, H. Schmutz, F. Merkt, and A. Wallraff, Phys. Rev. A {\bf 95}, 053855 (2017). \url{https://link.aps.org/doi/10.1103/PhysRevA.95.053855}

\bibitem{NatCommun:Hattermann2017} H. Hattermann, D. Bothner, L. Y. Ley, B. Ferdinand, D. Wiedmaier, L. S\'{a}rk\'{a}ny, R. Kleiner, D. Koelle, and J. Fort\'{a}gh, Nat. Commun. {\bf 8}, 2254 (2017). \url{https://doi.org/10.1038/s41467-017-02439-7}

\bibitem{Nature:Vijay2012} R. Vijay, C. Macklin, D. H. Slichter, S. J. Weber, K. W. Murch, R. Naik, A. N. Korotkov, and I. Siddiqi, Nature (London) {\bf 490}, 77 (2012). \url{http://dx.doi.org/10.1038/nature11505}

\end{thebibliography}
\end{document}